\documentclass[conference]{IEEEtran}
\IEEEoverridecommandlockouts
\usepackage{cite}
\usepackage{amsmath,amssymb,amsfonts}
\usepackage{algorithmic}
\usepackage{graphicx}
\usepackage{textcomp}
\usepackage{xcolor}
\usepackage{makecell}
\usepackage{hyperref}
\hypersetup{colorlinks=True, urlcolor=black}
\usepackage{multirow}
\usepackage{booktabs}
\def\BibTeX{{\rm B\kern-.05em{\sc i\kern-.025em b}\kern-.08em
    T\kern-.1667em\lower.7ex\hbox{E}\kern-.125emX}}

\makeatletter
\def\ps@IEEEtitlepagestyle{%
\def\@oddfoot{\mycopyrightnotice}%
\def\@evenfoot{}%
}
\def\mycopyrightnotice{%
{\footnotesize 978-1-6654-7189-3/22/\$31.00~\copyright~2022 IEEE\hfill}
\gdef\mycopyrightnotice{}
}

\begin{document}

\makeatletter
\def\ps@IEEEtitlepagestyle{%
\def\@oddfoot{\mycopyrightnotice}%
\def\@evenfoot{}%
}

\title{Singing Voice Synthesis with Vibrato Modeling and Latent Energy Representation}

\author{
\IEEEauthorblockN{Yingjie Song$^{1*}$, Wei Song$^2$, Wei Zhang$^2$, Zhengchen Zhang$^2$, Dan Zeng$^1$, Zhi Liu$^1$, Yang Yu$^{3}$}
\thanks{$^*$ Work was done during internship at JD AI Research.}
\IEEEauthorblockA{
\textit{$^1$ School of Communication and Information Engineering},
\textit{Shanghai University},
Shanghai, China \\
Email: \{jie0222, dzeng\}@shu.edu.cn, liuzhisjtu@163.com}

\IEEEauthorblockA{\textit{$^2$JD AI Research},
Beijing, China \\
Email: \{songwei11, Zhangzhengchen1\}@jd.com, wzhang.cu@gmail.com}

\IEEEauthorblockA{\textit{$^3$Shanghai Conservatory of Music},
Shanghai, China \\
Email: 15000401351@163.com}
}

\maketitle

\begin{abstract}
This paper proposes an expressive singing voice synthesis system by introducing explicit vibrato modeling and latent energy representation. 
Vibrato is essential to the naturalness of synthesized sound, due to the inherent characteristics of human singing. 
Hence, a deep learning-based vibrato model is introduced in this paper to control the vibrato's likeliness, rate, depth and phase in singing, where the vibrato likeliness represents the existence probability of vibrato and it would help improve the singing voice's naturalness.
Actually, there is no annotated label about vibrato likeliness in existing singing corpus. We adopt a novel vibrato likeliness labeling method to label the vibrato likeliness automatically. 
Meanwhile, the power spectrogram of audio contains rich information that can improve the expressiveness of singing. An autoencoder-based latent energy bottleneck feature is proposed for expressive singing voice synthesis. 
Experimental results on the open dataset NUS48E show that both the vibrato modeling and the latent energy representation could significantly improve the expressiveness of singing voice. The audio samples are shown
in the demo website\footnote{\url{https://mango321321.github.io/ExpressiveSing/}}.

\end{abstract}

\section{Introduction}
\label{sec:intro}
Singing voice synthesis (SVS) systems~\cite{zhuang2021litesing,hono2019singing,blaauw2020sequence,shi2021sequence,ren2020deepsinger,gao2020personalized,oura2010recent} generate singing voices from musical scores which contain music information such as lyrics, tempo, pitch, etc. SVS is similar to the text-to-speech (TTS) task~\cite{wang2017tacotron,ren2020fastspeech,song2020speaker,gupta2016physiological,jiang2008accurate,schnell2002text} in terms of generating speech from text. The difference is that SVS should reproduce emotional ebbs and flows beyond generating a natural-sounding voice~\cite{pang2004use,nwe2007exploring,luo2020singing}. The state-of-the-art singing voice synthesis systems already provide multiple solutions for the application, while synthesized voices lack expressiveness compared to real voices. 

Over the past few decades, researchers have made many efforts to enhance expressiveness. In~\cite{umbert2015expression}, Umbert \textit{et al.} presented the expressiveness is related to a set of voice parameters and classifies the expression control parameters into four categories including melody, dynamics, rhythm, and timbre. Rooted in~\cite{umbert2015expression}, many subsequent works for expressive singing voice synthesis focus on improvement about these control parameters, especially melody, dynamics \textit{a.k.a} pitch and energy. For example, to predict pitch feature better, Yi \textit{et al.}~\cite{yi2019singing} utilized deep autoregressive network to capture the dependencies among the consecutive acoustic features. Zhuang \textit{et al.}~\cite{zhuang2021litesing} separated the pitch feature from the acoustic feature to avoid the interdependence between these pitch features and the timbre features. Ren \textit{et al.}~\cite{ren2020fastspeech} introduced the pitch and energy information into the speech generation task and presented variance adaptors to make the generated audio expressive. In this article, we enhance expressiveness from these two aspects as well. 

Pitch conveys melody according to score and contains score-independent dynamic movements in a singing voice, such as vibrato. Because vibrato is not available in music score and it is unknown where the vibrato will appear, we introduced vibrato likeliness into pitch model to predict where the vibrato will occur. Vibrato likeliness represents the existence probability of vibrato but there is no annotated data in the existing dataset. Additional labeling is feasible but expensive. To address this problem, we propose a novel automatic labeling model called Vibrato Likeliness Labeling Network (VLLN), which is trained using the simulated data. 

Many studies show that energy is related to volume and prosody of the singing voice which affect expressiveness crucially. The energy prediction can be recast as the task of recovering the feature of power spectrogram.
In previous work~\cite{zhuang2021litesing,ren2020fastspeech,hantrakul2019fast}, the L2-Norm of the amplitude spectrogram is taken as energy information to promote expressiveness. However, this $1$-dimensional energy representation loses a lot of spectrogram information. In this work, an autoencoder-based subnetwork is used to learn a compact and representative latent energy representation from the input musical score information.

The main contributions of this paper are summarized as follows: 
\begin{enumerate}
\item An SVS system named ExpressiveSing is proposed which is adapted from an End-to-End TTS framework, and contains an accurate pitch model and a latent energy representation prediction module. 
\item The latent energy representation is proposed to utilize energy feature generating expressive singing voices.
\item A pitch model with accurate vibrato modeling is proposed by taking the vibrato likeliness into account. The vibrato likeliness is annotated by vibrato likeliness labeling method automatically. 

\end{enumerate}

The remainder of this paper is organized as follows: Section~\ref{sec:related work} describes the related work of pitch model and energy representation in SVS systems. Section~\ref{sec:PROPOSED Framework} introduces the proposed singing voice synthesis system in detail. The experimental setup is given in Section~\ref{sec:EXPERIMENTAL SETUP} and results are presented in Section~\ref{sec:EXPERIMENTAL RESULTS}. Conclusion follows in Section~\ref{sec:Conclusion}.

\section{Related Work}\label{sec:related work}
\subsection{Pitch Model in SVS}
The traditional pitch models include rule-based method, template-based method, unit concatenation method and HMM-based method. The rule-based method~\cite{berndtsson1996kth} designed rules for each singing style based on professional musical knowledge. It mainly depended on designer's experience, while template-based method proposed a simpler solution. Yukara \textit{et al.}~\cite{ikemiya2014transferring} chose typical examples from commercial recordings to form a library of parametrized templates. The unit concatenation method~\cite{umbert2013generating} used real contours extracted from singing data and concatenated them to generate pitch contours. Compared with the above mentioned methods, the HMM-based method~\cite{nose2015hmm} is more flexible. It generated pitch contours using hidden Markov models (HMMs). The disadvantage of this approach was that it was poor to retain the detailed information, resulting in the over-smoothing effect.    

Recently, data-driven approaches have been introduced into pitch modeling in SVS. They emphasized the vibrato during the pitch modeling process to further improve the expressiveness. Yi \textit{et al.}~\cite{yi2019singing} utilized the dependencies among the acoustic features of consecutive frames to help the pitch model understand vibrato. However, it still suffered the over-smoothing effect to some extent. Bonada \textit{et al.}~\cite{bonada2020hybrid} proposed a neural-parametric pitch model which solved the problem of over-smoothing effectively. It parametered the vibrato with phase, depth, and rate. Actually, there was noise mixed in vibrato component. Bonada \textit{et al.} directly constrained vibrato depth to be zero at note transitions, which was inelegant and could not eliminate noise completely. 

To solve the noise problem, we propose an additional parameter, namely vibrato likeliness, and automatically annotate the label of vibrato likeliness by the proposed Vibrato Likeliness Labeling Network. 

\subsection{Energy Representation in SVS}

Typically, the previous SVS systems implicitly modeled energy during synthesizing acoustic features such as~\cite{cao2021exploring,ren2020deepsinger,gu2021bytesing}. The TTS system, Fastspeech2~\cite{ren2020fastspeech} firstly proposed variance adaptor to model energy explicitly. It was proved to be beneficial for expressiveness. Concretely, the L2-norm of the amplitude of each short-time Fourier transform (STFT) frame was regarded as the energy and then the quantized energy is mapped to the embedding. In other words, the energy was compressed to 1-dimension vector firstly. Because the size of amplitude spectrogram is always thousands of dimensions, the operation of compression to 1-dimension vector would cause massive information loss. Following Fastspeech2, LiteSing \cite{zhuang2021litesing} makes the according modification in SVS system. The problem of information loss still existed. We propose an autoencoder-based latent spectrogram bottleneck feature to ease the information loss of compression. The experiment in Section~\ref{sec:EXPERIMENTAL RESULTS} shows that our proposed energy representation is superior to the one used in LiteSing and Fastspeech2.


\begin{figure}[b]
    \centering
    \includegraphics[width=0.48\textwidth]{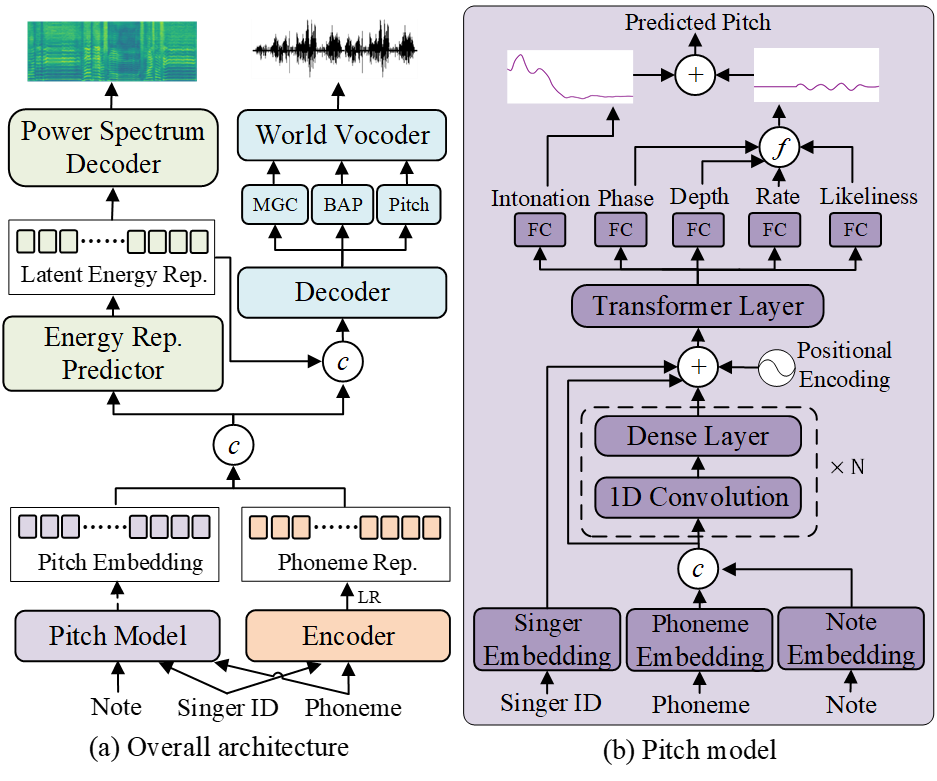}
    \caption{The overall architecture of ExpressiveSing, where `\textit{c}' denotes concatenation, and `+' denotes addition. `\textit{f}' refers to formula (\ref{equ_3}). `FC' is short for fully connected layer and `LR’ denotes length regulator. `Rep.’ stands for representation. The dashed arrow in subfigure~(a) indicates that the pitch model is trained separately. }
    \label{fig:Overall}
\end{figure}

\section{Proposed Framework}
\label{sec:PROPOSED Framework}

\subsection{Network Architecture}
The overall architecture of proposed ExpressiveSing is shown in Fig.~\ref{fig:Overall}(a). 
The network architecture of ExpressiveSing is a DIAN~\cite{song2021dian}-based network, which contains an encoder for extracting phoneme representation, and a decoder for the acoustic feature prediction. The encoder takes singer ID and phonemes at phoneme level as input and the following length regulator expands the generated phoneme representation to frame level. The decoder predicts acoustic features at frame level, \textit{i.e.} MGC, BAP, and pitch, which can be converted to audio by WORLD vocoder~\cite{morise2016world}. Compared to DIAN, ExpressiveSing designs pitch model, energy representation predictor and power spectrogram decoder additionally. The pitch model predicts human-like pitch contour from music notes. As for the input of pitch model, we use the note, phonemes at frame level and singer ID, so pitch model outputs pitch sequence at frame level. The energy representation predictor generates latent energy representation and the power spectrogram decoder reproduces the power spectrogram from the latent energy representation.


\subsection{Latent Energy Representation}
\label{ssec:Latent Representation of Energy}
An autoencoder-based network architecture is used in ExpressiveSing to predict the latent energy representation as shown in Fig.~\ref{fig:Overall}(a). The pitch embedding and phoneme representation are concatenated as the input of the energy representation predictor $P_e$. The energy representation predictor learns a compact and representative bottleneck feature $E \in R^ {T \times N}$ of power spectrogram, where $T$ is the frame count and $N$ denotes the dimension of the energy representation. The power spectrogram decoder $D_e$ composed of $1$D convolutions is trained to recover power spectrogram $S$ from the latent representation $E$. The power spectrogram decoder is jointly trained with acoustic feature prediction. The loss function for the power spectrogram decoder is shown as follows:
 \begin{equation*}
 \mathcal{L}(\textbf{S},\hat{\textbf{S}})=||\textbf{S}-D_e(\textbf{E})||_2
 \label{equ_1}
 \tag{1}
 \end{equation*}
where $\hat{S}$ denotes the predicted power spectrogram. 


\subsection{Pitch Model}
\label{ssec:Neural-parameter Pitch Model}
Pitch contour is decomposed into pitch intonation and vibrato~\cite{sundberg1996replicability,shonle1980pitch}. Intonation is responsible for singing in tune according to music note. Vibrato shows the skill of singers. 
We parameter the vibrato with phase, depth, rate, and the specially-designed likeliness. Vibrato depth and rate are the width and speed of the pitch fluctuation respectively~\cite{migita2010study,gu2014singing}.
Phase refers to the initial phase of vibrato, and likeliness stands for the possibility of the appearance of vibrato.
The training target of The vibrato depth, rate, and phase are extracted by Hilbert transformation.
The vibrato likeliness is annotated by the proposed Vibrato Likeliness Labeling Network, which is described in Section~\ref{subsec.vibrato.likeliness}.

Figure~\ref{fig:Overall}(b) illustrates the network architecture of the pitch model. The phoneme, music notes and singer ID are embedded as the input.

A Transformer~\cite{vaswani2017attention}-based network is used to predict the vibrato phase, depth, rate, likeliness and intonation at the frame level. As for the loss function, MSE loss is used in intonation, rate, depth, and phase prediction while the cross entropy loss is adopted for vibrato likeliness prediction. 


Following~\cite{yi2019singing}, to avoid out of tune, the predicted intonation ${\textbf{\textit{i}}}$ is post-processed by 
\begin{equation*}
  {\textbf{\textit{i}}}_{post}={\textbf{\textit{i}}}-\tilde{\textbf{\textit{i}}}+{\textbf{\textit{i}}}_{n}
 \label{equ_2}
 \tag{2}
\end{equation*}
where ${\textbf{\textit{i}}}_{n}$ means the music notes at frame level, and $\tilde{\textbf{\textit{i}}}$ is the smoothed curve of ${\textbf{\textit{i}}}$. A triangular window is used to convolve with ${\textbf{\textit{i}}}$ to get the smoothed result $\tilde{\textbf{\textit{i}}}$. 

For the predicted vibrato likeliness in pitch model, the average value $l_{mean}$ is calculated on each note.
If $l_{mean}$ is greater than the threshold $\epsilon$ ($\epsilon=0.5$ in our experiments), the note is considered to be a vibrato section, otherwise not. Finally, the vibrato ${v}_t~(t\in\{1,2,\cdots, T\})$ is synthesized by function:
\begin{equation*}
 {v}_t=\left\{
\begin{aligned}
&{l}_t \cdot{ e}_t \cdot \cos(\frac{ 2\pi {r}_t \cdot t}{f_s}+ {\phi}), &&{l_{mean} \geq \epsilon}\\ 
& 0,  &&{l_{mean} < \epsilon}
\end{aligned}
\right.
 \label{equ_3}
 \tag{3}
 \end{equation*}
where ${e}_t,~{r}_t,~{l}_t$ denote the $t$-th frame of predicted depth, rate, and likeliness of vibrato, respectively. $~{\phi}$ represents the predicted initial phase. ${f_s}$ stands for the sample rate.
Then the final output of pitch model is the sum of intonation ${\textbf{\textit{i}}}_{post}$ and predicted vibrato ${v}_t$.




\subsection{Vibrato Likeliness Labeling Network}
\label{subsec.vibrato.likeliness}
Training pitch model requires vibrato likeliness label, i.e., the ground truth of vibrato existence probability. 
However, no such label exists in existing singing corpus. To solve this problem, we propose a novel labeling method to get the ground truth of vibrato likeliness automatically. We randomly integrate vibrato to smoothed pitch contour to simulate the real pitch contour and then VLLN is trained on the simulated data. 

Concretely, the real pitch is extracted from the singing voice firstly and then the pitch contour is convolved with a triangular window to remove vibrato, resulting in a smoothed pitch contour. According to our observation, vibrato commonly appears in notes with a longer duration. Hence, we select notes with duration longer than one second to add simulated vibrato. In order to simulate vibrato on the selected music notes, we exploit a sinusoid curve in 6 Hz and vary its depth in range 0 to 2 in midi randomly and gradually. Then the synthesized vibrato is added to the smoothed pitch contour to generate the simulated data. Then frame-level labels of vibrato likeliness (0 or 1) can be obtained because we know which segment has vibrato. 

 \begin{figure}[b]
\centering
\includegraphics[width=0.25\textwidth]{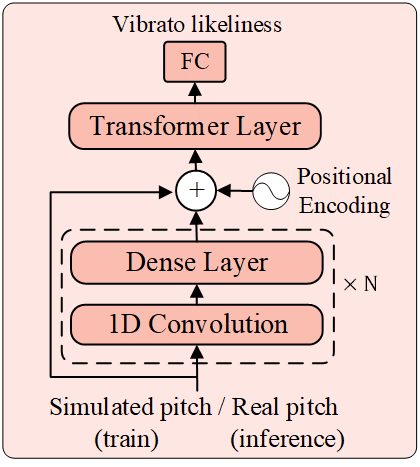}
\caption{The Vibrato Likeliness Labeling Network.}
\label{fig:VLLN}
\end{figure}

\begin{figure}[h]
\centering
\includegraphics[width=0.50\textwidth]{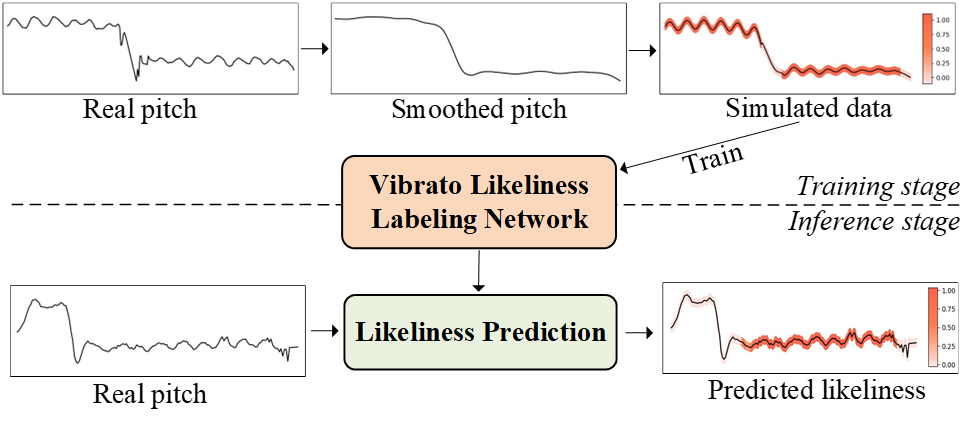}
\caption{The training and inference procedures of Vibrato Likeliness Labeling Network.}
\label{fig:VLLN_proceeding}
\end{figure}

The Vibrato Likeliness Labeling Network is shown in Fig.~\ref{fig:VLLN}, which is a Transformer-based network. Sigmoid function is used as activation function in the last layer. The cross entropy loss is adopted as the training objective for vibrato likeliness prediction. The details of training and inference procedures are shown in Fig.~\ref{fig:VLLN_proceeding}. During train stage, we employ the simulated data with known vibrato likeliness label to train VLLN. Afterwards, in inference stage, VLLN is used to label the vibrato likeliness for the real pitch of singing corpus. The color on the pitch contour represents vibrato likeliness. The darker the color, the higher the vibrato likeliness.

\section{Experimental Setup}
\label{sec:EXPERIMENTAL SETUP} 
\subsection{Dataset}
\label{ssec:Dataset}
NUS48E corpus~\cite{duan2013nus} is used to evaluate the proposed system, which consists of 48 popular English songs, and only 20 songs are unique. 
The lyrics are sung and read out by 12 singers (4 songs each singer). Only the singing data is used for the proposed model training, which contains 115 minutes recordings.

The training data are converted to 16kHz in our experiments. 
All the audios are split into segments shorter than 10 seconds and redundant silence of each utterance is removed. 
We annotated 36 songs with music scores manually because the NUS48E corpus does not contain music score labeling.
The pitch model is trained by 36 annotated songs while the acoustic model uses all 48 songs.
We keep $10\%$ as test set and use others for training. 
Pitch is extracted using the WORLD vocoder~\cite{morise2016world} and the 3$\sim$8Hz part extracted by the band-pass filter is considered to be vibrato, the hop size for acoustic feature extraction is set to 10 milliseconds in our experiments.
\subsection{Network Configuration and Implementation}
\label{ssec:Network configuration and training details}



We follow the network architecture configuration in DIAN~\cite{song2021dian}, and the energy representation predictor follows the variance adaptor setup in FastSpeech2~\cite{ren2020fastspeech} except that the output layer predicts a 256-dimension vector. 
The power spectrogram decoder contains a stack of 4 convolutional layers and a linear projection layer, the kernel size is 5 and channel size is 512 for each convolution layer, batch normalization~\cite{ioffe2015batch} and ReLU~\cite{agarap2018relu} activation is used for each convolution layer. The dimension of target power spectrogram is 1025. 
In training process of ExpressiveSing, we use real pitch as input. In inference stage, the predicted pitch from pitch model is taken as input.
Table \ref{tabel_1} shows the dimension of features used in ExpressiveSing.


The proposed pitch model follows the duration model network configuration in~\cite{song2021dian} and has the following modifications: (1) music notes are also taken as input as well as the phoneme input; (2) the pitch model contains 5 output layers, which are used to predict intonation, phase, depth, rate and likeliness.
The loss weight of intonation prediction is 1, and the loss weight of vibrato rate, depth, phase, and likeliness are all 1e-2. 

The Vibrato Likeliness Labeling Network's configuration is the same as pitch model, except that there is no singer embedding in Vibrato Likeliness Labeling Network and only contains 1 output layer.


Adam optimizer~\cite{kingma2014adam} with batch size 32 is used for optimization with $\beta_1$ = 0.9 and $\beta_2$ = 0.999.
ExpressiveSing is trained with learning rate that starts from 1e-3. The learning rate exponentially decays to 1e-5 in the first 50K steps, then the training continued with a constant learning rate of 1e-5 until 60K steps.
The pitch model and Vibrato Likeliness Labeling Network are trained with Noam\cite{vaswani2017attention} learning rate schedule with a peak learning rate of 1e-4, with a total of 40k steps for each model. 


\begin{table}[]
\centering  
\caption{The dimension of features used in ExpressiveSing.}  
\label{tabel_1}
\begin{tabular}{l|l}
\toprule[1pt]
\multicolumn{1}{c|}{Feature} & \multicolumn{1}{c}{Dimension} \\ \hline
Phone Representation                & \makecell[c]{512 }                          \\
Pitch Embedding                     & \makecell[c]{32 }                           \\ 
Latent Energy Representation   & \makecell[c]{256}                           \\ 
Mel-Generalized Cepstral            &\makecell[c]{60 }                          \\ 
Band Aperiodicity                   & \makecell[c]{4}                            \\ 
Pitch                               & \makecell[c]{1}                             \\ \toprule[1pt]
\end{tabular}
\vspace{-3mm}
\end{table} 

\section{Experimental Results}
\label{sec:EXPERIMENTAL RESULTS}

\begin{table}[]
\centering  
\caption{Mean opinion scores (MOS) of different systems. Pronunciation accuracy (Pronun acc.), sound quality (Quality), and naturalness are evaluated separately.}
\label{table_2}
\begin{tabular}{l|l l l l}
\toprule[1pt]
\multicolumn{1}{c|}{System}         & \multicolumn{1}{c}{Pronun acc.}    & Quality                     & Naturalness  \\ \hline
Baseline                            & \makecell[c]{3.120}                & \makecell[c]{2.373}         & \makecell[c]{2.460}           \\ 
USVC\cite{nachmani2019unsupervised} & \makecell[c]{3.160}                & \makecell[c]{2.993}         & \makecell[c]{2.707}            \\ 
WGANSing\cite{chandna2019wgansing}  & \makecell[c]{3.367}                & \makecell[c]{2.960}         & \makecell[c]{3.133}           \\ 

UCSVC\cite{polyak2020unsupervised} & \makecell[c]{\textbf{3.900}}                 & \makecell[c]{\textbf{3.547}}         & \makecell[c]{\textbf{3.780}}              \\ 
ExpressiveSing                     &  \makecell[c]{\underline{3.780}}                & \makecell[c]{\underline{3.413}}         & \makecell[c]{\underline{3.467}}        \\
\toprule[1pt]
\end{tabular}
\vspace{-3mm}
\end{table}

\subsection{Overall Performance}
\label{ssec:Overall performance}
The system performance is evaluated using subjective mean opinion scores (MOS). A total of $15$ subjects participated in the evaluation. The listeners were asked to mark the singing samples from three aspects: pronunciation accuracy, sound quality and naturalness. The scope of the score is from 1 (extremely bad) to 5 (extremely good). Five systems are compared, and ten samples generated by each system are prepared for the evaluation. The baseline in our experiment is built using the network structure proposed in DIAN~\cite{song2021dian}.
The difference between baseline and ExpressiveSing is that the baseline system has no pitch model nor latent energy representation module. The proposed ExpressiveSing is also compared with the state-of-the-art systems including two singing voice conversion systems USVC~\cite{nachmani2019unsupervised} and UCSVC~\cite{polyak2020unsupervised}, as well as a singing synthesis system WGANSing~\cite{chandna2019wgansing}. These systems are trained on the same NUS48E dataset. It is notable that the singing voice conversion systems have real human singing pitch as input, which would make the generated voices more natural. The samples of USVC\footnote{\url{https://enk100.github.io/Unsupervised_Singing_Voice_Conversion/}} and UCSVC\footnote{\url{https://singing-conversion.github.io/}} are downloaded from their demo web pages. 
We implemented the WGANSing method, and trained it on the same training set as ExpressiveSing. Because WGANSing uses the real pitch originally, to ensure fairness, WGANSing is conditioned with the predicted pitch from our pitch model in this experiment. 
The evaluation results are shown in Table~\ref{table_2}. 
We can see that ExpressiveSing (ES) outperforms WGANSing, USVC and the baseline system. And ES obtains comparable scores with UCSVC, even though UCSVC uses real pitch as input and exploits additional speech datasets for training.
 
\subsection{Ablation Study}
\begin{figure}[h]
\centering
\includegraphics[width=0.45\textwidth]{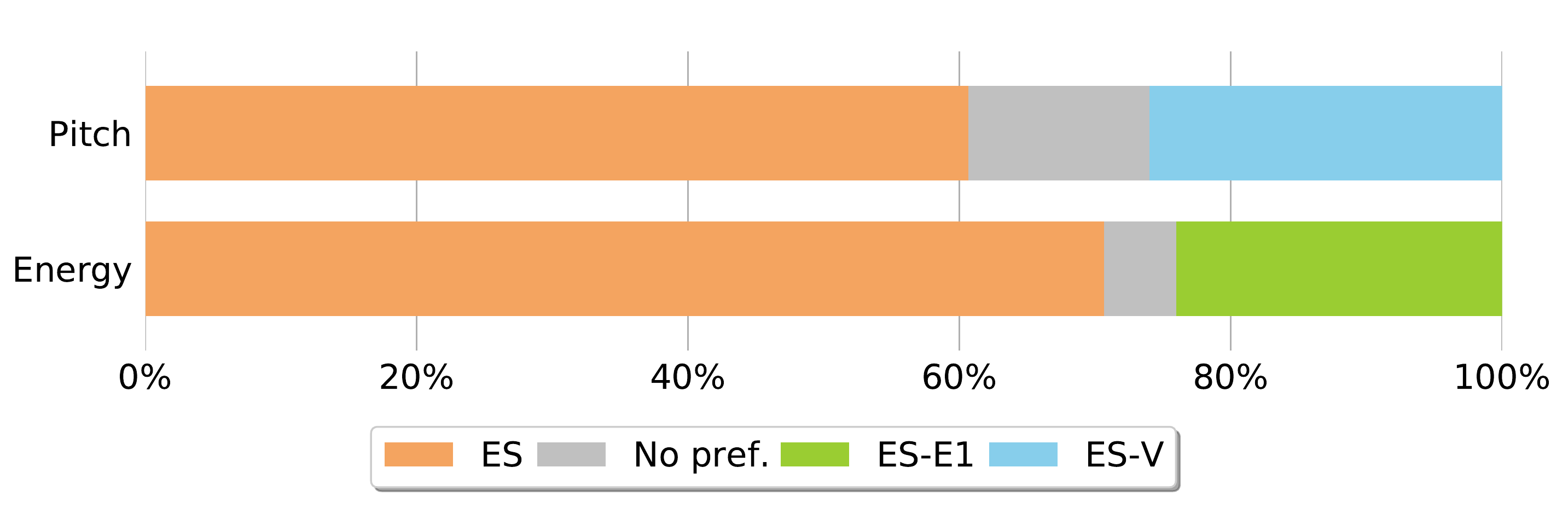}
\caption{A/B preference evaluation results of ablation study. `No pref.' denotes no preference for the systems.   }
\label{fig:abtest}
\end{figure}

We conduct ablation studies to evaluate the effects of the proposed pitch model, and the latent energy representation. Both subjective and objective evaluations are performed. 
For subjective evaluation, the listeners are instructed to do the A/B preference tests to compare the naturalness of the singing samples generated by different systems. Two pairs of systems are compared. One is the comparison between ExpressiveSing (ES) and ExpressiveSing without vibrato likeliness (ES-V), where ES-V uses the pitch model without the parameter of vibrato likeliness. The other is the comparison between ExpressiveSing (ES) and ExpressiveSing without latent energy representation (ES-E1). ES-E1 compressed energy to 1-dimension vector firstly and then a lookup table is used to convert this single valued energy to embedding representation, which is the same as LiteSing and Fastspeech2. The evaluation results are shown in Fig.~\ref{fig:abtest}. 
One can find that the listeners strongly prefer the ES model to either ES-V or ES-E1, which proves the effectiveness of the proposed pitch model and latent energy representation. 


We further evaluate the pitch model with objective metrics including F0 Root Mean Square Error (RMSE) and Correlation (CORR). The results are shown in Table~\ref{table_3}. The synthesized F0 sequences are compared with the human singers' F0 sequences and the score's pitch notations, which are noted as `Natural' and 'Score' respectively. 
The more natural the generated singing is, the lower F0 RMSE and higher CORR are achieved. 
We can find that ES achieved lower F0 RMSE and higher CORR than ES-V. To compare the proposed pitch model with the state-of-the-art methods, we further implemented the pitch model proposed in~\cite{bonada2020hybrid}, which is named Parametric F0. The F0 RMSE and CORR of Parametric F0 are also presented in Table~\ref{table_3}. It can be seen that ES achieved better performance than Parametric F0.



\begin{table}[]
\centering 
\caption{Objective evaluation of systems with different pitch models. ExpressiveSing (ES), ExpressiveSing without vibrato likeliness (ES-V), and the model proposed in~\cite{bonada2020hybrid} are compared. }
\label{table_3}
\begin{tabular}{l|l|l|l|l}
\hline\toprule[1pt]
\multicolumn{1}{c|}{\multirow{2}{*}{System}} & \multicolumn{2}{c|}{F0 RMSE (Hz)} & \multicolumn{2}{{c}}{F0 CORR} \\ \cline{2-5} 
\multicolumn{1}{c|}{}                & Natural          & Score          & Natural      & Score      \\ \hline
Parametric F0~\cite{bonada2020hybrid} & \makecell[c]{22.759} &\makecell[c]{23.764} &\makecell[c]{0.973}&\makecell[c]{0.980} \\
ES-V                     & \makecell[c]{21.837}&\makecell[c]{14.459}& \makecell[c]{0.977}&\makecell[c]{0.993} \\ 
ES                       & \makecell[c]{\textbf{20.900}}& \makecell[c]{\textbf{12.633}}& \makecell[c]{\textbf{0.978}}& \makecell[c]{\textbf{0.996}}   \\ 
 \hline\toprule[1pt]
\end{tabular}
\end{table}

\begin{figure}[h]
\centering
\includegraphics[width=0.45\textwidth]{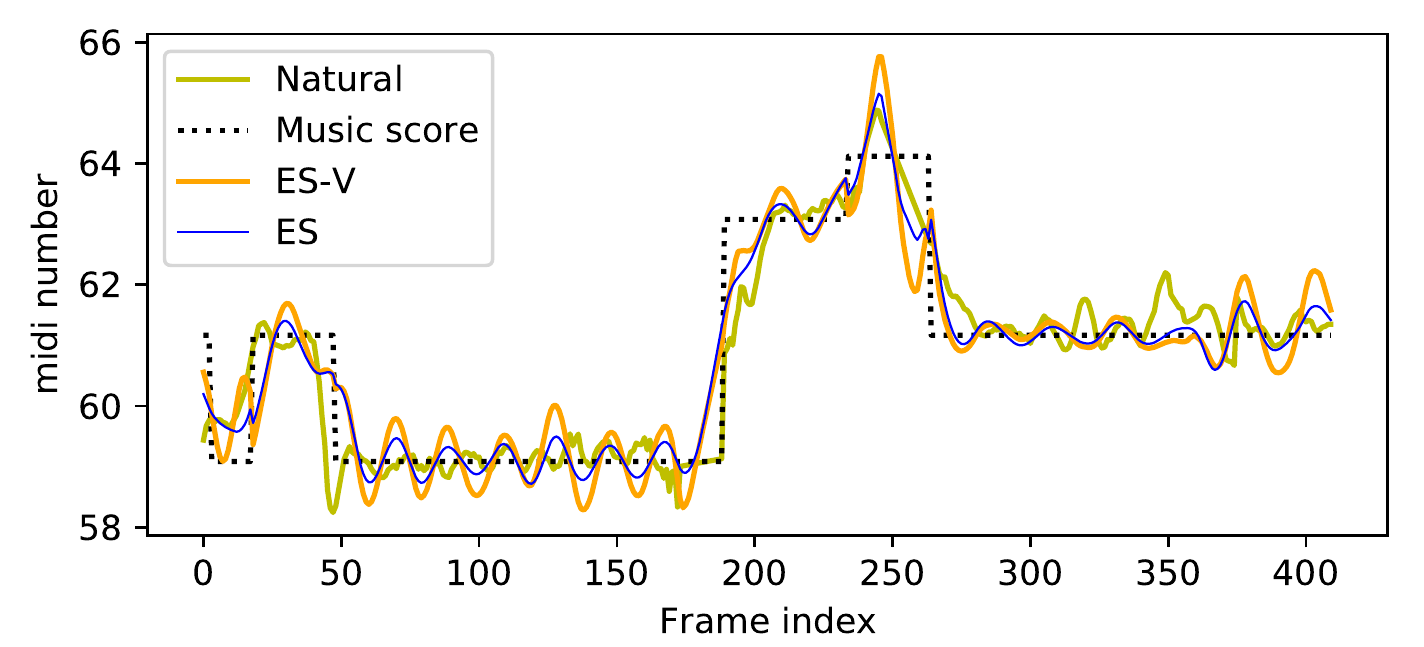}
\caption{An example of pitch contour generated by ES and ES-V, compared with human singing voices (Natural) and the note of score (Music Note).}
\label{fig:pitch}
\end{figure}

An example is illustrated in Fig.~\ref{fig:pitch} to demonstrate the effect of the pitch model. Compared to ES-V, ES generates fewer fluctuations at the positions of note changing with the help of vibrato likeliness prediction. Hence, the noise in the non-vibrato section is avoided, and the generated sound becomes more steady. 


Mel-cepstral distortion (MCD) metric is used to verify if the proposed latent energy representation is helpful.
Three systems, ES, ES-E1, and ES-w/o, are evaluated, where ES-w/o is the EpressiveSing model without energy information. 
The MCD scores are 6.600, 6.787, and 6.800 respectively. ES obtained the lowest MCD, which means the generated singing voice is more similar to human singing. 
The result is consistent with the A/B test results. ES-w/o obtained the highest MCD, which proved that the energy feature is helpful to expressive singing voice synthesis.
\section{Conclusion}
\label{sec:Conclusion}


According to the inherent characteristics of human singing, accurate vibrato modeling and latent energy representation are introduced into the singing voice synthesis system in this paper.
The MOS results show that the proposed system can synthesize expressive singing voice.
The A/B test results show that both the proposed vibrato modeling and latent energy representation could improve the naturalness.
The objective metrics of MCD, F0 RMSE and CORR are consistent with the A/B test results.
Both the subjective and objective results verify the effectiveness of vibrato modeling of pitch and latent energy representation.
\vfill\pagebreak

\footnotesize
\bibliographystyle{IEEEtran}
\bibliography{IEEEexample}

\begin{thebibliography}{10}
\providecommand{\url}[1]{#1}
\csname url@samestyle\endcsname
\providecommand{\newblock}{\relax}
\providecommand{\bibinfo}[2]{#2}
\providecommand{\BIBentrySTDinterwordspacing}{\spaceskip=0pt\relax}
\providecommand{\BIBentryALTinterwordstretchfactor}{4}
\providecommand{\BIBentryALTinterwordspacing}{\spaceskip=\fontdimen2\font plus
\BIBentryALTinterwordstretchfactor\fontdimen3\font minus
  \fontdimen4\font\relax}
\providecommand{\BIBforeignlanguage}[2]{{%
\expandafter\ifx\csname l@#1\endcsname\relax
\typeout{** WARNING: IEEEtran.bst: No hyphenation pattern has been}%
\typeout{** loaded for the language `#1'. Using the pattern for}%
\typeout{** the default language instead.}%
\else
\language=\csname l@#1\endcsname
\fi
#2}}
\providecommand{\BIBdecl}{\relax}
\BIBdecl

\bibitem{zhuang2021litesing}
X.~Zhuang, T.~Jiang, S.-Y. Chou, B.~Wu, P.~Hu, and S.~Lui, ``Litesing:
  {T}owards {F}ast, {L}ightweight and {E}xpressive {S}inging {V}oice
  {S}ynthesis,'' in \emph{ICASSP 2021-2021 IEEE International Conference on
  Acoustics, Speech and Signal Processing (ICASSP)}.\hskip 1em plus 0.5em minus
  0.4em\relax IEEE, 2021, pp. 7078--7082.

\bibitem{hono2019singing}
Y.~Hono, K.~Hashimoto, K.~Oura, Y.~Nankaku, and K.~Tokuda, ``Singing {V}oice
  {S}ynthesis {B}ased on {G}enerative {A}dversarial {N}etworks,'' in
  \emph{ICASSP 2019-2019 IEEE International Conference on Acoustics, Speech and
  Signal Processing (ICASSP)}.\hskip 1em plus 0.5em minus 0.4em\relax IEEE,
  2019, pp. 6955--6959.

\bibitem{blaauw2020sequence}
M.~Blaauw and J.~Bonada, ``Sequence-to-{S}equence {S}inging {S}ynthesis {U}sing
  the {F}eed-{F}orward {T}ransformer,'' in \emph{ICASSP 2020-2020 IEEE
  International Conference on Acoustics, Speech and Signal Processing
  (ICASSP)}.\hskip 1em plus 0.5em minus 0.4em\relax IEEE, 2020, pp. 7229--7233.

\bibitem{shi2021sequence}
J.~Shi, S.~Guo, N.~Huo, Y.~Zhang, and Q.~Jin, ``Sequence-{T}o-{S}equence
  {S}inging {V}oice {S}ynthesis {W}ith {P}erceptual {E}ntropy {L}oss,'' in
  \emph{ICASSP 2021-2021 IEEE International Conference on Acoustics, Speech and
  Signal Processing (ICASSP)}.\hskip 1em plus 0.5em minus 0.4em\relax IEEE,
  2021, pp. 76--80.

\bibitem{ren2020deepsinger}
Y.~Ren, X.~Tan, T.~Qin, J.~Luan, Z.~Zhao, and T.-Y. Liu, ``Deep{S}inger:
  {S}inging {V}oice {S}ynthesis with {D}ata {M}ined {F}rom the {W}eb,'' in
  \emph{Proceedings of the 26th ACM SIGKDD International Conference on
  Knowledge Discovery \& Data Mining}, 2020, pp. 1979--1989.

\bibitem{gao2020personalized}
X.~Gao, X.~Tian, Y.~Zhou, R.~K. Das, and H.~Li, ``Personalized {S}inging
  {V}oice {G}eneration {U}sing {W}ave{RNN},'' in \emph{Proc. Odyssey 2020 The
  Speaker and Language Recognition Workshop}, 2020, pp. 252--258.

\bibitem{oura2010recent}
K.~Oura, A.~Mase, T.~Yamada, S.~Muto, Y.~Nankaku, and K.~Tokuda, ``Recent
  development of the hmm-based singing voice synthesis system—sinsy,'' in
  \emph{Seventh ISCA Workshop on Speech Synthesis}, 2010.

\bibitem{wang2017tacotron}
Y.~Wang, R.~Skerry-Ryan, D.~Stanton, Y.~Wu, R.~J. Weiss, N.~Jaitly, Z.~Yang,
  Y.~Xiao, Z.~Chen, S.~Bengio \emph{et~al.}, ``Tacotron: {T}owards
  {E}nd-to-{E}nd {S}peech {S}ynthesis,'' \emph{arXiv preprint
  arXiv:1703.10135}, 2017.

\bibitem{ren2020fastspeech}
Y.~Ren, C.~Hu, X.~Tan, T.~Qin, S.~Zhao, Z.~Zhao, and T.-Y. Liu, ``Fast{S}peech
  2: Fast and {H}igh-{Q}uality {E}nd-to-{E}nd {T}ext to {S}peech,'' \emph{arXiv
  preprint arXiv:2006.04558}, 2020.

\bibitem{song2020speaker}
E.~Song, J.-S. Kim, K.~Byun, and H.-G. Kang, ``Speaker-adaptive neural vocoders
  for parametric speech synthesis systems,'' in \emph{2020 IEEE 22nd
  International Workshop on Multimedia Signal Processing (MMSP)}.\hskip 1em
  plus 0.5em minus 0.4em\relax IEEE, 2020, pp. 1--5.

\bibitem{gupta2016physiological}
R.~Gupta and T.~H. Falk, ``Physiological quality-of-experience assessment of
  text-to-speech systems,'' in \emph{2016 IEEE 18th International Workshop on
  Multimedia Signal Processing (MMSP)}.\hskip 1em plus 0.5em minus 0.4em\relax
  IEEE, 2016, pp. 1--2.

\bibitem{jiang2008accurate}
D.~Jiang, I.~Ravyse, H.~Sahli, and Y.~Zhang, ``Accurate visual speech synthesis
  based on diviseme unit selection and concatenation,'' in \emph{2008 IEEE 10th
  Workshop on Multimedia Signal Processing}.\hskip 1em plus 0.5em minus
  0.4em\relax IEEE, 2008, pp. 906--909.

\bibitem{schnell2002text}
M.~Schnell, M.~Kustner, O.~Jokisch, and R.~Hoffmann, ``Text-to-speech for
  low-resource systems,'' in \emph{2002 IEEE Workshop on Multimedia Signal
  Processing.}\hskip 1em plus 0.5em minus 0.4em\relax IEEE, 2002, pp. 259--262.

\bibitem{pang2004use}
H.-S. Pang, ``On the use of the maximum likelihood estimation for analysis of
  vibrato tones,'' \emph{Applied Acoustics}, vol.~65, no.~1, pp. 101--107,
  2004.

\bibitem{nwe2007exploring}
T.~L. Nwe and H.~Li, ``Exploring {V}ibrato-{M}otivated {A}coustic {F}eatures
  for {S}inger {I}dentification,'' \emph{IEEE Transactions on Audio, Speech,
  and Language Processing}, vol.~15, no.~2, pp. 519--530, 2007.

\bibitem{luo2020singing}
Y.-J. Luo, C.-C. Hsu, K.~Agres, and D.~Herremans, ``Singing voice conversion
  with disentangled representations of singer and vocal technique using
  variational autoencoders,'' in \emph{ICASSP 2020-2020 IEEE International
  Conference on Acoustics, Speech and Signal Processing (ICASSP)}.\hskip 1em
  plus 0.5em minus 0.4em\relax IEEE, 2020, pp. 3277--3281.

\bibitem{umbert2015expression}
M.~Umbert, J.~Bonada, M.~Goto, T.~Nakano, and J.~Sundberg, ``Expression
  {C}ontrol in {S}inging {V}oice {S}ynthesis: {F}eatures, approaches,
  evaluation, and challenges,'' \emph{IEEE Signal Processing Magazine},
  vol.~32, no.~6, pp. 55--73, 2015.

\bibitem{yi2019singing}
Y.-H. Yi, Y.~Ai, Z.-H. Ling, and L.-R. Dai, ``Singing {V}oice {S}ynthesis
  {U}sing {D}eep {A}utoregressive {N}eural {N}etworks for {A}coustic
  {M}odeling,'' \emph{arXiv preprint arXiv:1906.08977}, 2019.

\bibitem{hantrakul2019fast}
L.~Hantrakul, J.~H. Engel, A.~Roberts, and C.~Gu, ``Fast and {F}lexible
  {N}eural {A}udio {S}ynthesis,'' in \emph{ISMIR}, 2019.

\bibitem{berndtsson1996kth}
G.~Berndtsson, ``The kth rule system for singing synthesis,'' \emph{Computer
  Music Journal}, vol.~20, no.~1, pp. 76--91, 1996.

\bibitem{ikemiya2014transferring}
Y.~Ikemiya, K.~Itoyama, and H.~G. Okuno, ``Transferring vocal expression of f0
  contour using singing voice synthesizer,'' in \emph{International Conference
  on Industrial, Engineering and Other Applications of Applied Intelligent
  Systems}.\hskip 1em plus 0.5em minus 0.4em\relax Springer, 2014, pp.
  250--259.

\bibitem{umbert2013generating}
M.~Umbert, J.~Bonada, and M.~Blaauw, ``Generating singing voice expression
  contours based on unit selection,'' in \emph{Proc. SMAC}, 2013.

\bibitem{nose2015hmm}
T.~Nose, M.~Kanemoto, T.~Koriyama, and T.~Kobayashi, ``Hmm-based expressive
  singing voice synthesis with singing style control and robust pitch
  modeling,'' \emph{Computer Speech \& Language}, vol.~34, no.~1, pp. 308--322,
  2015.

\bibitem{bonada2020hybrid}
J.~Bonada and M.~Blaauw, ``Hybrid {N}eural-{P}arametric {F}0 {M}odel for
  {S}inging {S}ynthesis,'' in \emph{ICASSP 2020-2020 IEEE International
  Conference on Acoustics, Speech and Signal Processing (ICASSP)}.\hskip 1em
  plus 0.5em minus 0.4em\relax IEEE, 2020, pp. 7244--7248.

\bibitem{cao2021exploring}
Y.~Cao, S.~Liu, S.~Kang, N.~Hu, P.~Liu, X.~Liu, D.~Su, D.~Yu, and H.~Meng,
  ``Exploring {C}ross-lingual {S}inging {V}oice {S}ynthesis {U}sing {S}peech
  {D}ata,'' in \emph{2021 12th International Symposium on Chinese Spoken
  Language Processing (ISCSLP)}.\hskip 1em plus 0.5em minus 0.4em\relax IEEE,
  2021, pp. 1--5.

\bibitem{gu2021bytesing}
Y.~Gu, X.~Yin, Y.~Rao, Y.~Wan, B.~Tang, Y.~Zhang, J.~Chen, Y.~Wang, and Z.~Ma,
  ``Byte{S}ing: A {C}hinese {S}inging {V}oice {S}ynthesis {S}ystem {U}sing
  {D}uration {A}llocated {E}ncoder-{D}ecoder {A}coustic {M}odels and
  {W}ave{RNN} {V}ocoders,'' in \emph{2021 12th International Symposium on
  Chinese Spoken Language Processing (ISCSLP)}.\hskip 1em plus 0.5em minus
  0.4em\relax IEEE, 2021, pp. 1--5.

\bibitem{song2021dian}
W.~Song, X.~Yuan, Z.~Zhang, C.~Zhang, Y.~Wu, X.~He, and B.~Zhou, ``Dian:
  {D}uration {I}nformed {A}uto-{R}egressive {N}etwork for {V}oice {C}loning,''
  in \emph{ICASSP 2021-2021 IEEE International Conference on Acoustics, Speech
  and Signal Processing (ICASSP)}.\hskip 1em plus 0.5em minus 0.4em\relax IEEE,
  2021, pp. 8598--8602.

\bibitem{morise2016world}
M.~Morise, F.~Yokomori, and K.~Ozawa, ``World: {A} {V}ocoder-{B}ased
  {H}igh-{Q}uality {S}peech {S}ynthesis {S}ystem for {R}eal-{T}ime
  {A}pplications,'' \emph{IEICE TRANSACTIONS on Information and Systems},
  vol.~99, no.~7, pp. 1877--1884, 2016.

\bibitem{sundberg1996replicability}
J.~Sundberg, E.~Prame, and J.~Iwarsson, ``Replicability and accuracy of pitch
  patterns in professional singers,'' \emph{Vocal fold physiology, controlling
  complexity and chaos}, pp. 291--306, 1996.

\bibitem{shonle1980pitch}
J.~I. Shonle and K.~E. Horan, ``The pitch of vibrato tones,'' \emph{The Journal
  of the Acoustical Society of America}, vol.~67, no.~1, pp. 246--252, 1980.

\bibitem{migita2010study}
N.~Migita, M.~Morise, and T.~Nishiura, ``A {S}tudy of {V}ibrato {F}eatures to
  {C}ontrol {S}inging {V}oices,'' \emph{Proc. ICA}, pp. 23--27, 2010.

\bibitem{gu2014singing}
H.-Y. Gu and Z.-F. Lin, ``Singing-voice {S}ynthesis {U}sing {ANN}
  {V}ibrato-parameter {M}odels.'' \emph{J. Inf. Sci. Eng.}, vol.~30, no.~2, pp.
  425--442, 2014.

\bibitem{vaswani2017attention}
A.~Vaswani, N.~Shazeer, N.~Parmar, J.~Uszkoreit, L.~Jones, A.~N. Gomez,
  {\L}.~Kaiser, and I.~Polosukhin, ``Attention {I}s {A}ll {Y}ou {N}eed,''
  \emph{Advances in neural information processing systems}, vol.~30, 2017.

\bibitem{duan2013nus}
Z.~Duan, H.~Fang, B.~Li, K.~C. Sim, and Y.~Wang, ``The nus sung and spoken
  lyrics corpus: {A} quantitative comparison of singing and speech,'' in
  \emph{2013 Asia-Pacific Signal and Information Processing Association Annual
  Summit and Conference}.\hskip 1em plus 0.5em minus 0.4em\relax IEEE, 2013,
  pp. 1--9.

\bibitem{ioffe2015batch}
S.~Ioffe and C.~Szegedy, ``Batch {N}ormalization: {A}ccelerating {D}eep
  {N}etwork {T}raining by {R}educing {I}nternal {C}ovariate {S}hift,'' in
  \emph{International conference on machine learning}.\hskip 1em plus 0.5em
  minus 0.4em\relax PMLR, 2015, pp. 448--456.

\bibitem{agarap2018relu}
A.~F. Agarap, ``Deep {L}earning using {R}ectified {L}inear {U}nits
  ({R}e{L}u),'' \emph{arXiv preprint arXiv:1803.08375}, 2018.

\bibitem{kingma2014adam}
D.~P. Kingma and J.~Ba, ``Adam: {A} {M}ethod for {S}tochastic {O}ptimization,''
  \emph{arXiv preprint arXiv:1412.6980}, 2014.

\bibitem{nachmani2019unsupervised}
E.~Nachmani and L.~Wolf, ``Unsupervised {S}inging {V}oice {C}onversion,'' in
  \emph{Interspeech 2019, 20th Annual Conference of the International
  Speech}.\hskip 1em plus 0.5em minus 0.4em\relax {ISCA}, 2019, pp. 2583--2587.

\bibitem{chandna2019wgansing}
P.~Chandna, M.~Blaauw, J.~Bonada, and E.~G{\'o}mez, ``Wgansing: {A}
  {M}ulti-{V}oice {S}inging {V}oice {S}ynthesizer {B}ased on the
  {W}asserstein-{GAN},'' in \emph{2019 27th European Signal Processing
  Conference (EUSIPCO)}.\hskip 1em plus 0.5em minus 0.4em\relax IEEE, 2019, pp.
  1--5.

\bibitem{polyak2020unsupervised}
A.~Polyak, L.~Wolf, Y.~Adi, and Y.~Taigman, ``Unsupervised {C}ross-{D}omain
  {S}inging {V}oice {C}onversion,'' in \emph{Interspeech 2020, 21st Annual
  Conference of the International Speech}.\hskip 1em plus 0.5em minus
  0.4em\relax {ISCA}, 2020, pp. 801--805.

\end{thebibliography}
\vspace{12pt}
\color{red}

\end{document}